\documentclass[12pt]{article}  %fuer beidseitige Version
\usepackage{graphicx}
\usepackage[a4paper,textwidth=490.0pt,textheight=703.1pt]{geometry}
\usepackage{color}
\usepackage{cite}
\usepackage[rflt]{floatflt}

\newcommand{\ie}{i.e.,\ }
\renewcommand{\H}[2]{\textnormal{H}_{#1}(#2)}

\usepackage{array}

\usepackage[english]{babel}
\usepackage[latin1]{inputenc}
\usepackage[T1]{fontenc}
\usepackage{ae}

\usepackage{url}

\usepackage{amsmath, amsthm, amssymb}
\newtheorem{thm}{Theorem}[section]

\newtheorem{definition}[thm]{Definition}

 \newcommand{\HA}{{\rm H }}

\renewcommand{\H}[2]{\textnormal{H}_{#1}\hspace{-0.2em}\left(#2\right)}
\newcommand{\R}{\mathbb R}

\newcommand{\Li}{{\rm Li}}
\newcommand{\Mvec}{{\rm {\bf  M}}}

\newcommand{\ep}{\varepsilon}

\usepackage{rotating}
\newcommand{\shuffle}{\, \raisebox{1.2ex}[0mm][0mm]{\rotatebox{270}{$\exists$}} \,}
\usepackage{algorithm}
\usepackage{algorithmicx}
\usepackage{algpseudocode}
\usepackage{graphicx}
\newcounter{mmacnt}
\def\restartmma{\setcounter{mmacnt}{0}}
\restartmma \catcode`|=\active
\def|#1|{\mathrm{#1}}
\catcode`|=12
\newenvironment{mma}{
 \par\smallskip
 \catcode`|=\active
 \parskip=0pt\parindent=0pt % locally
 \small
 \def\In##1\\{%
   \def\linebreak{\hfill\break\null\qquad}%
   \refstepcounter{mmacnt}
   \hangindent=2.5em\hangafter=0
   \leavevmode
   \llap{\tiny\sffamily In[\arabic{mmacnt}]:=\kern.5em}%
   \mathversion{bold}\footnotesize$\displaystyle##1$\normalsize
   \mathversion{normal}\par
 }%
 \def\Print##1\\{%
   \def\linebreak{\hfill\break}%
   \hangindent=2.5em\hangafter=0
   \leavevmode ##1\par}%
 \def\Out##1\\{%
   \def\linebreak{$\hfill\break\null\hfill$}%
   \kern\abovedisplayskip\par
   \hangindent=2.5em\hangafter=0
   \leavevmode
   \llap{\tiny\sffamily Out[\arabic{mmacnt}]=\kern.5em}
   \footnotesize$\displaystyle##1$\normalsize\hfill\null\par
   \kern\belowdisplayskip
 }%
 \def\Warning##1##2\\{%
   \def\linebreak{\hfill\break}%
   \hangindent=2.5em\hangafter=0
   \leavevmode
   {\scriptsize##1 : ##2}\par}%
}{%
 \par\smallskip
}

\usepackage{color}

\newenvironment{fshaded}{%
\MakeFramed {\FrameRestore}
}%
{\endMakeFramed}

\usepackage{tikz}
\usetikzlibrary{matrix}

\allowdisplaybreaks[4]

\usepackage{hyperref}

\begin{document}
\setlength{\baselineskip}{0.515cm}
\sloppy
\thispagestyle{empty}
\begin{flushleft}
DESY 21--031
\hfill 
%{\tt arXiv:2103.xxxxx [hep-th]}
\\
DO--TH 21/05 \\
RISC-Linz Report Series No. 21-05\\
SAGEX-21-05 \\
March 2021\\
\end{flushleft}

\mbox{}
\vspace*{\fill}
\begin{center}

{\Large\bf Iterated integrals over letters induced by quadratic forms}

\vspace{4cm}
\large
J.~Ablinger$^a$,
J.~Bl\"umlein$^b$
and C.~Schneider$^a$ 

\vspace{1.5cm}
\normalsize
{\it $^a$~Research Institute for Symbolic Computation (RISC),\\
Johannes Kepler University, Altenbergerstra\ss{}e 69,
A-4040, Linz, Austria}
\\

\vspace*{3mm}
{\it  $^b$ Deutsches Elektronen-Synchrotron, DESY,}\\
{\it  Platanenallee 6, D-15738 Zeuthen, Germany}\\

%%\today

\end{center}
\normalsize
\vspace{\fill}
\begin{abstract}
\noindent 
An automated treatment of iterated integrals based on letters induced by real-valued quadratic forms 
and Kummer--Poincar\'e letters is presented. These quantities emerge in analytic single and multi--scale 
Feynman diagram calculations. To compactify representations, one wishes to apply general properties of 
these quantities in computer-algebraic implementations. We provide the reduction to basis representations, 
expansions, analytic continuation and numerical evaluation of these quantities.
\end{abstract}

\vspace*{\fill}
\noindent
\numberwithin{equation}{section}
%%%%%%%%%%%%%%%%%%%%%%%%%%%%%%%%%%%%%%%%%%%%%%%%%%%%%%%%%%%%%%%%%%%%%%%
\newpage

%%%%%%%%%%%%%%%%%%%%%%%%%%%%%%%%%%%%%%%%%%%%%%%%%%%%%%%%%%%%%%%%%%%%%%%
\section{Introduction}
\label{sec:1}
%%%%%%%%%%%%%%%%%%%%%%%%%%%%%%%%%%%%%%%%%%%%%%%%%%%%%%%%%%%%%%%%%%%%%%%

\vspace*{1mm}
\noindent
In analytic calculations of single and multi--scale Feynman integrals different principal structures have 
been revealed in particular during the last 30 years.\footnote{For a survey see \cite{Blumlein:2018cms}.} 
Beyond the multiple zeta values \cite{Blumlein:2009cf} and other special numbers for zero--scale quantities, 
there are the spaces of harmonic sums \cite{Vermaseren:1998uu,Blumlein:1998if}, harmonic polylogarithms 
\cite{Remiddi:1999ew}, generalized harmonic sums \cite{Moch:2001zr,Ablinger:2013cf} and Kummer--Poincar\'{e} 
iterated integrals \cite{KUMMER,Moch:2001zr,Ablinger:2013cf}, cyclotomic harmonic sums and iterated integrals 
\cite{Ablinger:2011te}, finite and infinite binomial sums and inverse binomial sums and the associated 
root--letter integrals \cite{Ablinger:2014bra,BINOM}, and iterative non--iterative integrals 
\cite{Ablinger:2017bjx}, 
including those containing complete elliptic integrals \cite{ELLIPTIC,Ablinger:2017bjx}. This list is 
expected still to extend in analytic calculations at even higher loops and for more contributing scales in 
the future.

In decomposing Feynman parameter representations in the general case, cf.~e.g.~\cite{Bogner:2010kv},
often real polynomials of higher degree have to be factored. According to the fundamental theorem of 
algebra  \cite{ARGAND} this leads
to either linear and quadratic factors in real representations or to linear complex--valued factors with conjugated 
pairs. Real representations have often advantages in calculations. This is the main reason to extend the class
of Kummer--Poincar\'{e} iterated integrals based on the alphabet 
%--------------------------------------------------------------------------------------------------
\begin{eqnarray}
\mathfrak{A}_{KP} = \left\{\left. \frac{1}{x-c_i} \right| c_i \in \mathbb{C} \right\}
\end{eqnarray}
%--------------------------------------------------------------------------------------------------
into\footnote{Here the condition in (\ref{eq:AR}) implies that the polynomial $x^2+b_ix+c_i$ is irreducible 
over $\mathbb{R}$.}
%--------------------------------------------------------------------------------------------------
\begin{eqnarray}
\label{eq:AR}
\mathfrak{A}_{R} = \left\{\left. \frac{1}{x-a_i}, \frac{1}{x^2 + b_i x + c_i}, \frac{x}{x^2 + b_i x + c_i} 
\right| a_i, b_i, c_i \in \mathbb{R},~ 4 c_i \geq b_i \right\}.
\end{eqnarray}
%--------------------------------------------------------------------------------------------------
There is an overlap with the cyclotomic iterated integrals \cite{Ablinger:2011te} w.r.t. to 
the letters $1/\Phi_k(x), k = 3, 4, 6$. The associated iterative integrals are given by
%--------------------------------------------------------------------------------------------------
\begin{eqnarray}
\HA_{b,\vec{a}}(z) = \int_0^z dx f_b(x) \HA_{\vec{a}}(x),~~\HA_\emptyset = 1,~~f_l(x) \in \mathfrak{A}_R.
\end{eqnarray}
%--------------------------------------------------------------------------------------------------
Iterated integrals do also obey the differential property
%--------------------------------------------------------------------------------------------------
\begin{eqnarray}
\label{eq:DIFF}
\frac{d}{dz} \HA_{b,\vec{a}}(z) = f_b(z) \HA_{\vec{a}}(z), 
\end{eqnarray}
%--------------------------------------------------------------------------------------------------
which is instrumental for building the corresponding algebra, to be closed under differentiation. 
Sometimes more general iterated integrals are used, cf.~e.g.~\cite{Lee:2021iid}, with letters outside 
$\mathfrak{A}_R$. As we will show below this class of integrals can be cast into the class generated by
$\mathfrak{A}_R$.

Using the alphabet 
(\ref{eq:AR}) has the advantage to stay inside a real representation in calculating a real quantity. 
Complex decompositions \cite{Vollinga:2004sn} request the thorough observation of the pairing of complex 
conjugated letters. Both approaches can handle poles inside the integration region.

We will device an algorithm to transform the mentioned formal iterated integrals into real ones, also referring 
to one simple main variable. This has the advantage that the corresponding results can be iterated over in further 
integrations, which will be necessary for the use in a higher order calculation.

Using iterative integrals in the description of physical quantities it is required to give them a clear definition.
In some cases it is possible that a first definition has singularities in sub-integrals, which have to be dealt with
to obtain a measurable quantity. Furthermore, the real and imaginary parts of the respective integrals have to be 
separated from the beginning, because they have a different physical meaning and it then allows to deal with real 
integrals only.
One has also to observe
that certain transformations in the main argument may effect the position of cuts chosen. In the case of singularities
of the real integrals we will apply Cauchy's principal value for definiteness, as it is also the case in amplitudes 
referring to the K\"all\'{e}n--Lehmann representation \cite{CAUCHY}. 

In Section~\ref{sec:2} we will describe the different operations  for the iterated integrals induced by quadratic forms
in the package {\tt HarmonicSums} 
\cite{HARMSU,Vermaseren:1998uu,Blumlein:1998if,Remiddi:1999ew,Ablinger:2011te,
Ablinger:2013cf,Ablinger:2014bra,AB1,AB2} 
and 
provide test examples. Section~\ref{sec:3} deals with integrals of a recent physical application \cite{Lee:2021iid} 
which we reconsider in the present formalism, and Section~\ref{sec:4} contains the conclusion.
%%%%%%%%%%%%%%%%%%%%%%%%%%%%%%%%%%%%%%%%%%%%%%%%%%%%%%%%%%%%%%%%%%%%%%%
\section{Operations for the iterated integrals}
\label{sec:2}
%%%%%%%%%%%%%%%%%%%%%%%%%%%%%%%%%%%%%%%%%%%%%%%%%%%%%%%%%%%%%%%%%%%%%%%

\vspace*{1mm}
\noindent
In the following we describe a series of operations which allow to deal with iterated integrals containing letters 
of the alphabet $\mathfrak{A}_R$.

The statement
%---------------------------------------------------------------------------------------------------------------
\begin{eqnarray}
{\tt
QL[\{
\{\{a_1, b_1, c_1\}, d_1\}, 
\{\{a_2, b_2, c_2\}, d_2\}, 
\{\{a_3, b_3, c_3\}, d_3\}\}, {\it z}]} 
\end{eqnarray}
%---------------------------------------------------------------------------------------------------------------
represents the integral
%---------------------------------------------------------------------------------------------------------------
\begin{eqnarray}
& & \HA_{
((a_1,b_1,c_1),d_1),
((a_2,b_2,c_2),d_2),
((a_3,b_3,c_3),d_3)}(z) = 
\nonumber\\ && \hspace*{5cm}
\int_0^z  \hspace*{-2mm}
\int_0^{t_1} \hspace*{-2mm}
\int_0^{t_2} \frac{t_1^{d_1}t_2^{d_2}t_3^{d_3} dt_1 dt_2 dt_3}
{(a_1 + b_1 t_1 + c_1 t_1^2) 
 (a_2 + b_2 t_2 + c_2 t_2^2) 
 (a_3 + b_3 t_3 + c_3 t_3^2)},
\nonumber\\ && \hspace*{5cm}
~\text{with}~d_i \in \{0,1\}, a_i, b_i, c_i \in \mathbb{R}, 
\end{eqnarray}
%---------------------------------------------------------------------------------------------------------------
covering a number of iterated letters out of $\mathfrak{A}_R$. The command {\tt ToHarmonicSumsIntegrate}
reveals the integral structure in explicite form. One may convert these integrals into {\tt GL}--functions, 
cf.~\cite{Ablinger:2014bra}, by {\tt QLToGL} and {\tt GL}--functions with letters out of $\mathfrak{A}_R$ to 
{\tt QL}--functions by {\tt GLToQL}. 

It is allowed that the Kummer--Poincar\'{e} letters in $\mathfrak{A}_{KP}$ have poles in the integration region.
The iterative integral is then defined taking Cauchy's principal value. However, the quadratic denominators are assumed 
to not factorize in real numbers.

The numerical evaluation of {\tt QL}--functions is performed as in the following example
\begin{eqnarray}
r &=& {\tt ToHarmonicSumsIntegrate\big[
  QL \left[\{\{\{1, 1, 2\}, 3\}, \{\{1, -1, 1\}, 1\}, \{\{-1, 1, 1\}, 
     1\}\}, \tfrac{1}{2}\right],} 
\nonumber\\ 
& & 
{\tt NIntegrate \rightarrow   True \big] // ReleaseHold}
\\
r &\approx& {\tt -0.0000649218}.
\end{eqnarray}
%-----------------------------------------------------------------------------------------------------------
Here not all letters corresponding to quadratic forms are yet in the standard form. There is no 
singularity, however, in the integration region since $(\sqrt{5}-1)/2 > 1/2$.

To be able to deal with properly defined letters, the mapping {\tt QLToStandardForm} is used. A typical example is
%-----------------------------------------------------------------------------------------------------------
\begin{eqnarray}
{\tt QL[\{\{\{1, 1, 2\}, 2\}\}, {\it z}]} &=& \int_0^z dt \frac{t^2}{1+t+2 t^2}
\\
{\tt QLToStandardForm[QL[\{\{\{1, 1, 2\}, 2\}\}, {\it z}]]} &\Rightarrow& \frac{z}{2} 
-\frac{1}{4}\int_0^z dt \frac{1}{\tfrac{1}{2} + \tfrac{t}{2} +  t^2}
-\frac{1}{4}\int_0^z dt \frac{t}{\tfrac{1}{2} + \tfrac{t}{2} +  t^2}.
\nonumber\\
\end{eqnarray}
%-----------------------------------------------------------------------------------------------------------

The shuffle operation is
%-----------------------------------------------------------------------------------------------------------
\begin{eqnarray}
\HA_{a_1,...a_m}(z) \cdot \HA_{b_1,...b_n}(z) =   \HA_{a_1,...a_n}(z) \shuffle \HA_{b_1,...b_n}(z) = 
\sum_{c_i \in \{{a_1,...a_m} \shuffle {b_1,...b_n}\}} \HA_{c_i}(z),  
\label{eq:shuf}
\end{eqnarray}
%-----------------------------------------------------------------------------------------------------------
where all combinations of the two index sets are allowed, which preserve the ordering in these two sets.
Here $\HA$ labels a general iterated integral. 
The corresponding command is {\tt LinearHExpand}.

One obtains
%-----------------------------------------------------------------------------------------------------------
\begin{eqnarray}
&& {\tt 
LinearHExpand[QL\left[\{\{\{\tfrac{1}{2}, \tfrac{1}{2}, 1\}, 0\}\}, \tfrac{1}{2}\right]^2 
QL\left[\{\{\{\tfrac{1}{2}, \tfrac{1}{2}, 2\}, 0\}\}, \tfrac{1}{2}\right]]
} \Rightarrow 
\nonumber\\ && 
\hspace*{5.3cm}
{\tt 2 \bigl\{
QL\left[\{\{\{\tfrac{1}{2}, \tfrac{1}{2}, 1\}, 0\}, \{\{\tfrac{1}{2}, \tfrac{1}{2}, 1\}, 0\}, 
\{\{\tfrac{1}{2},\tfrac{1}{2}, 2\}, 0\}\}, \tfrac{1}{2}\right]}  
\nonumber\\ && 
\hspace*{5.3cm}
{\tt +
QL\left[\{\{\{\tfrac{1}{2}, \tfrac{1}{2}, 1\}, 0\}, \{\{\tfrac{1}{2}, \tfrac{1}{2}, 2\}, 0\}, 
\{\{\tfrac{1}{2},\tfrac{1}{2}, 1\}, 0\}\}, \tfrac{1}{2}\right]}  
\nonumber\\ && 
\hspace*{5.3cm}
{\tt +
QL\left[\{\{\{\tfrac{1}{2}, \tfrac{1}{2}, 2\}, 0\}, \{\{\tfrac{1}{2}, \tfrac{1}{2}, 1\}, 0\}, 
\{\{\tfrac{1}{2},\tfrac{1}{2}, 1\}, 0\}\}, \tfrac{1}{2}\right]  
}\bigr\}.
\nonumber\\
\label{eq:shuf1}
\end{eqnarray}
%-----------------------------------------------------------------------------------------------------------
The algebraic reduction w.r.t. shuffle relations of a given expression is performed by the command {\tt 
ReduceToQLBasis}, which will also transform the shuffled expression (\ref{eq:shuf1}) into the corresponding product 
expression.

Often one would like to remove trailing indices or leading indices of {\tt QL}--functions. This is done by the commands
{\tt RemoveTrailing0, RemoveTrailingIndex} or {\tt RemoveLeading1, RemoveLeadingIndex}. Here {\tt Trailing0} refers to 
the letter $1/x$ and {\tt Leading1} to $1/(1-x)$. Examples are
%-----------------------------------------------------------------------------------------------------------
\begin{eqnarray}
&& {\tt RemoveTrailingIndex[
 QL[\{\{\{1, 1, 1\}, 0\}, \{\{-1, 1, 0\}, 0\}, \{\{1, 1, 1\}, 0\}, \{\{1, 1, 1\}, 0\}\},
   {\it z}],} 
\nonumber\\ & & 
\{\{1, 1, 1\}, 0\}]} =
{\tt \frac{1}{2} QL[\{\{\{1, 1, 1\}, 0\}\}, {\it z}]^2 QL[\{\{{1, 1, 1\}, 0\}, \{\{-1, 1, 0\}, 0\}\}, 
   {\it z}]} 
\nonumber\\ & & 
{\tt - 2 QL[\{\{\{1, 1, 1\}, 0\}\}, 
   {\it z}] QL[\{\{\{1, 1, 1\}, 0\}, \{\{1, 1, 1\}, 0\}, \{\{-1, 1, 0\}, 0\}\}, {\it z}]} 
\nonumber\\ & & 
{\tt + 
 3 QL[\{\{\{1, 1, 1\}, 0\}, \{\{1, 1, 1\}, 0\}, \{\{1, 1, 1\}, 0\}, \{\{-1, 1, 0\}, 
     0\}\}, {\it z}]},
\\
%-----
&& {\tt RemoveLeadingIndex[
QL[\{\{\{1, 1, 1\}, 0\}, \{\{-1, 1, 0\}, 0\}, \{\{1, 1, 1\}, 0\}\}, {\it z}], \{\{1, 1, 1\},0\}]} 
\nonumber\\ & & 
= {\tt QL[\{\{\{1, 1, 1\}, 0\}\}, {\it z}] QL[\{\{\{-1, 1, 0\}, 
0\}, \{\{1, 1, 1\}, 0\}\}, {\it z}]} 
\nonumber\\ & & 
{\tt - 2 QL[\{\{\{-1, 1, 0\}, 0\}, \{\{1, 1, 1\}, 0\}, \{\{1, 1, 1\}, 0\}\}, {\it z}]}.
\end{eqnarray}
%-----------------------------------------------------------------------------------------------------------

The series expansion of the {\tt QL}--function about $z=0$ is obtained by the command {\tt QLSeries}, e.g.
%-----------------------------------------------------------------------------------------------------------
\begin{eqnarray}
{\tt QLSeries\left[QL\left[\{\{\{1, 1, 1\}, 1\}, \{\{3, 1, 1\}, 0\}, \{\{\tfrac{1}{2}, -1, 1\}, 1\}\}\},   {\it 
z}\right], 
{\it z}, 10 \right]}  
\nonumber\\ &&  \hspace*{-7cm} =
\frac{z^5}{45} - \frac{z^6}{216} - \frac{11 z^7}{1260} 
+ \frac{13 z^8}{1620} - \frac{7 z^9}{972} - \frac{
 4793 z^{10}}{680400}
\end{eqnarray}
%-----------------------------------------------------------------------------------------------------------
or
%-----------------------------------------------------------------------------------------------------------
\begin{eqnarray}
{\tt 
QLSeries\left[QL\left[\{\{\{1, 1, 1\}, 1\}, \{\{3, 1, 1\}, 0\}, \{\{0, 1, 0\}, 0\}\}, 
  {\it z}\right], {\it z}, 5\right]}
\nonumber\\ &&  \hspace*{-7cm} =
-\frac{4 z^3}{27} + \frac{11 z^4}{96} - \frac{5 z^5}{972}
+ \left[\frac{z^3}{9} - \frac{7 z^4}{72} + \frac{z^5}{162} \right]
\ln(z).
\end{eqnarray}
%-----------------------------------------------------------------------------------------------------------
Often a change in the argument of the {\tt QL}--functions is desirable. The following transformations are implemented
and are carried out by the command {\tt TransformQL}
%-----------------------------------------------------------------------------------------------------------
\begin{eqnarray}
k \cdot z \rightarrow z,~~k \in \mathbb{Q},~~~~
1 - z \rightarrow z,~~~~
\frac{1}{z} \rightarrow z,~~~~
\frac{1-z}{1+z} \rightarrow z.
\end{eqnarray}
%-----------------------------------------------------------------------------------------------------------
Examples are
%-----------------------------------------------------------------------------------------------------------
\begin{eqnarray}
{\tt TransformQL[QL[\{\{\{1, 1, 1\}, 1\}, \{\{2, 1, 1\}, 0\}\}, 2 {\it z}], {\it z}]} &\Rightarrow& \hspace*{-2mm}
{\tt \frac{1}{2} QL\left[\{\{\{\tfrac{1}{4}, \tfrac{1}{2}, 1\}, 1\}, \{\{\tfrac{1}{2}, \tfrac{1}{2}, 1\}, 0\}\}, {\it 
x}\right]},
\nonumber\\
\end{eqnarray}
%-----------------------------------------------------------------------------------------------------------
%-----------------------------------------------------------------------------------------------------------
\begin{eqnarray}
{\tt TransformQL[QL[\{\{\{1, 1, 1\}, 1\}, \{\{2, 1, 1\}, 0\}\}, 1 - {\it z}], {\it z}]} 
\nonumber\\
&& 
\hspace*{-8cm}
\Rightarrow {\tt QL[\{\{\{2, 1, 1\}, 0\}\}, 1] (
-QL[\{\{\{3, -3, 1\}, 0\}\}, {\it z}] + QL[\{\{\{3, -3, 1\}, 1\}\}, {\it z}])} 
\nonumber\\
&& \hspace*{-8cm}
{\tt
+ 
 QL[\{\{\{1, 1, 1\}, 1\}, \{\{2, 1, 1\}, 0\}\}, 1] + 
 QL[\{\{\{3, -3, 1\}, 0\}, \{\{4, -3, 1\}, 0\}\}, {\it z}]}  
\nonumber\\
&& 
\hspace*{-8cm}
-{\tt
 QL[\{\{\{3, -3, 1\}, 1\}, \{\{4, -3, 1\}, 0\}\}, {\it z}]},
\end{eqnarray}
%-----------------------------------------------------------------------------------------------------------
%-----------------------------------------------------------------------------------------------------------
\begin{eqnarray}
\hspace*{-2cm}
{\tt TransformQL\left[QL\left[\{\{\{1, 1, 1\}, 1\}, \{\{2, 1, 1\}, 0\}\}, \frac{1}{\it z}\right], {\it z}\right]} 
&\Rightarrow& 
\nonumber\\
&& \hspace*{-10cm} 
\Biggl[
-{\tt QL[\{\{\{0,1,0\},0\}\},{\it z}]
     -QL[\{\{\{1,1,1\},0\}\},1]
     +QL[\{\{\{1,1,1\},0\}\},{\it z}]}
\nonumber\\ &&
\hspace*{-10cm} 
{\tt -QL[\{\{\{1,1,1\},1\}\},1] 
+QL[\{\{\{1,1,1\},1\}\},{\it z}]\Biggr] + \frac{1}{2}
\Biggl[QL\left[\{\{\{\tfrac{1}{2}\tfrac{1}{2},1\},0\}\},1\right]}
\nonumber\\ &&
\hspace*{-10cm} 
{\tt +2 
QL[\{\{\{2,1,1\},0\}\},1]) 
-QL\left[\{\{\{0,1,0\},0\},\{\{\tfrac{1}{2},\tfrac{1}{2},1\},0\}\},1\right]}
\nonumber\\ &&
\hspace*{-10cm} 
{\tt
+QL\left[\{\{\{0,1,0\},0\},\{\{\tfrac{1}{2},\tfrac{1}{2},1\},0\}\},{\it z}\right]
+QL\left[\{\{\{1,1,1\},0\},\{\{\tfrac{1}{2},\tfrac{1}{2},1\},0\}\},1\right]}
\nonumber\\ &&
\hspace*{-10cm} 
{\tt
-QL\left[\{\{\{1,1,1\},0\},\{\{\tfrac{1}{2},\tfrac{1}{2},1\},0\}\},{\it z}\right]
+QL\left[\{\{\{1,1,1\},1\},\{\{\tfrac{1}{2},\tfrac{1}{2},1\},0\}\},1\right]}
\nonumber\\ &&
\hspace*{-10cm} 
{\tt -QL\left[\{\{\{1,1,1\},1\},\{\{\tfrac{1}{2},\tfrac{1}{2},1\},0\}\},{\it z}\right]} {\tt +2 
QL[\{\{\{1,1,1\},1\},\{\{2,1,1\},0\}\},1])}
\Biggr],
\end{eqnarray}
%-----------------------------------------------------------------------------------------------------------
%-----------------------------------------------------------------------------------------------------------
\begin{eqnarray}
{\tt TransformQL\left[QL\left[\{\{\{1, 1, 1\}, 1\}, \{\{2, 1, 1\}, 0\}\}, \tfrac{1-{\it z}}{1+{\it z}}\right], {\it 
z}\right]} &\Rightarrow& 
\nonumber\\
&& \hspace*{-8cm}
{\tt QL[\{\{\{2,1,1\},0\}\},1] 
(QL[\{\{\{1,1,0\},0\}\},1]
-QL[\{\{\{1,1,0\},0\}\},{\it z}]})
\nonumber\\
&& \hspace*{-8cm}
   {\tt
   -QL[\{\{\{3,0,1\},0\}\},1]
+QL[\{\{\{3,0,1\},0\}\},{\it z}]
-QL[\{\{\{3,0,1\},1\}\},1]}
\nonumber\\
&& \hspace*{-8cm}
{\tt +QL[\{\{\{3,0,1\},1\}\},{\it z}])
-QL[\{\{\{1,1,0\},0\},\{\{2,1,1\},0\}\},1]}
\nonumber\\
&& \hspace*{-8cm}
{\tt +QL[\{\{\{1,1,0\},0\},\{\{2,1,1\},0\}\},{\it z}]
+QL[\{\{\{3,0,1\},0\},\{\{2,1,1\},0\}\},1]}
\nonumber\\
&& \hspace*{-8cm}
{\tt -QL[\{\{\{3,0,1\},0\},\{\{2,1,1\},0\}\},{\it z}]
+QL[\{\{\{3,0,1\},1\},\{\{2,1,1\},0\}\},1]}
\nonumber\\
&& \hspace*{-8cm}
{\tt
-QL[\{\{\{3,0,1\},1\},\{\{2,1,1\},0\}\},{\it z}]
}.
\label{eq:MOB1}
\end{eqnarray}
%-----------------------------------------------------------------------------------------------------------
In these transformations new special numbers are occurring beyond the multiple zeta values. In particular one 
obtains the values
%-----------------------------------------------------------------------------------------------------------
\begin{eqnarray}
{\tt QL[\{\{\{3, 0, 1\}, 1\}\}, 1]} &=& \ln(2) - \frac{1}{2} \ln(3)
\\ 
{\tt QL\left[\{\{\{\tfrac{1}{2}, \tfrac{1}{2}, 1\}, 0\}\}, 1\right]} &=& \frac{4}{\sqrt{7}} 
\arctan\left(\frac{\sqrt{7}}{3}\right)
\\
{\tt QL[\{\{\{2, 1, 1\}, 0\}\}, 1]} &=& \frac{2}{\sqrt{7}} \arctan\left(\frac{\sqrt{7}}{5}\right).
\end{eqnarray}
%-----------------------------------------------------------------------------------------------------------
More involved constants are dilogarithms of arguments e.g. of the type $ \sqrt{3} (1+ i \sqrt{3})/(\sqrt{7}+\sqrt{3})$, 
etc. and further special constants.

It is furthermore useful to transform {\tt QL}--functions in a representation in which they have a convergent
Taylor series representation, which is also applied in their numerical evaluation. This is provided by the command
{\tt QLToConvergentRegion}. An example is
%-----------------------------------------------------------------------------------------------------------
\begin{eqnarray}
&& {\tt QLToConvergentRegion[QL[\{\{\{1, 1, 0\}, 0\}, \{\{-1, 1, 0\}, 0\}, \{\{2, 1, 1\}, 1\}\}, 4]]}
\nonumber\\ 
&=& {\tt QL[\{\{\{0,1,0\},0\}\},3] (QL[\{\{\{1,1,0\},0\}\},1] QL[\{\{\{2,1,1\},1\}\},1]+QL[\{\{\{2,1,1\},1\}\},1]} 
\nonumber\\ && 
{\tt QL[\{\{\{4,1,0\},0\}\},1])+QL[\{\{\{2,1,1\},1\}\},1](-QL[\{\{\{0,1,0\},0\},\{\{1,1,0\},0\}\},1]}
\nonumber\\ && 
{\tt -QL[\{\{\{2,1,0\},0\},\{\{4,1,0\},0\}\},1])+QL[\{\{\{1,1,0\},0\}\},1] (-QL[\{\{\{2,1,0\},0\}\},1]} 
\nonumber\\ && 
{\tt QL[\{\{\{2,1,1\},1\}\},1]-QL[\{\{\{2,1,1\},1\},\{\{-1,1,0\},0\}\},1])+QL[\{\{\{4,1,0\},0\}\},1]}
\nonumber\\ && 
\times {\tt (\tfrac{1}{2} QL[
\{\{\{0,1,0\},0\},\{\{1,\tfrac{3}{2},1\},0\}\},1]+QL[\{\{\{0,1,0\},0\},\{\{1,\tfrac{3}{2},1\},1\}\},1]}
\nonumber\\ && 
               {\tt -QL[\{\{\{2,1,1\},1\},\{\{-1,1,0\},0\}\},1])
               +\tfrac{1}{2} QL[\{\{\{1,\tfrac{3}{2},1\},0\}\},1]} 
\nonumber\\ && 
\times {\tt QL[\{\{\{4,1,0\},0\},\{\{2,1,0\},0\}\},1]}
{\tt + QL[\{\{\{4,1,0\},0\},\{\{2,1,0\},0\}\},1]}
\nonumber\\ && 
{\tt \times QL[\{\{\{1,\tfrac{3}{2},1\},1\}\},1]} 
{\tt +QL[\{\{\{1,1,0\},0\},\{\{-1,1,0\},0\},\{\{2,1,1\},1\}\},1]}
\nonumber\\ && 
{\tt +\tfrac{1}{2} 
QL[\{\{\{1,1,0\},0\},\{\{0,1,0\},0\},\{\{1,\tfrac{3}{2},1\},0\}\},1]}
\nonumber\\ &&
{\tt +QL[\{\{\{1,1,0\},0\},\{\{0,1,0\},0\},\{\{1,\tfrac{3}{2},1\},
1\}\},1]}
\nonumber\\ && 
{\tt +3 QL[\{\{\{4,1,0\},0\},\{\{2,1,0\},0\},\{\{14,7,1\},0\}\},1]}
\nonumber\\ && 
{\tt +QL[
\{\{\{4,1,0\},0\},\{\{2,1,0\},0\},\{\{14,7,1\},1\}\},1]}.  
\end{eqnarray}
%-----------------------------------------------------------------------------------------------------------

For removing poles in the integration domain we rely on the formula
\begin{align}\label{RemovePole1}
 \H{a_1,a_2,\ldots,a_m}z&=\sum_{i=0}^m\H{b_1,b_2,\ldots,b_i}{z-p}\H{a_{i+1},a_{i+2},\ldots,a_m}{p}
\end{align}
where $f_{b_j}(x):=f_{a_j}(x+p)$.

The following example illustrates the use of this formula to remove poles. 
Consider $\H{((0,1,0),0),((-1,1,0),0),((1,0,1),0)}{z}$. It has a pole at $1$ due to the letter 
%-----------------------------------------------------------------------------------------------------------
\begin{eqnarray}
\frac{1}{-1 + x} = ((-1,1,0),0).
\end{eqnarray}
%-----------------------------------------------------------------------------------------------------------
In order to remove this pole we use \eqref{RemovePole1} which yields 
%-----------------------------------------------------------------------------------------------------------
\begin{align*}
 \H{((0,1,0),0),((-1,1,0),0),((1,0,1),0)}{z}=&\H{((0,1,0),0),((-1,1,0),0),((1,0,1),0)}{1}\\
 &+\H{((1,1,0),0)}{z-1}\H{((-1,1,0),0),((1,0,1),0)}{1}\\
 &+\H{((1,1,0),0),((0,1,0),0)}{z-1}\H{((1,0,1),0)}{1}\\
 &+\H{((1,1,0),0),((0,1,0),0),((2,2,1),0)}{z-1}.
\end{align*}
%-----------------------------------------------------------------------------------------------------------
Remove trailing zeros, \ie the letter $((0,1,0),0)$ from the functions with argument $x-1$, yields
%-----------------------------------------------------------------------------------------------------------
\begin{align*}
&\H{((0,1,0),0),((-1,1,0),0),((1,0,1),0)}{1}+\H{((1,1,0),0)}{z-1}\H{((-1,1,0),0),((1,0,1),0)}{1}\\
 &+\left[\H{((1,1,0),0)}{z-1}\H{((0,1,0),0)}{z-1}-\H{((0,1,0),0),((1,1,0),0)}{z-1}\right]\H{((1,0,1),0)}{1}\\
 &+\H{((1,1,0),0),((0,1,0),0),((2,2,1),0)}{z-1}.
\end{align*}
%-----------------------------------------------------------------------------------------------------------------
Now we replace 
%-----------------------------------------------------------------------------------------------------------------
\begin{eqnarray}
\H{((0,1,0),0)}{z-1} =  \H{((0,1,0),0)}{\frac{z-1}{p}}-\H{((-1,1,0),0)}{1}+v\cdot i\pi
\end{eqnarray}
%-----------------------------------------------------------------------------------------------------------------
and finally we remove leading ones, \ie the letter $((-1,1,0),0)$, which results in
%-----------------------------------------------------------------------------------------------------------------
\begin{align*}
\H{((0,1,0),0),((-1,1,0),0),((1,0,1),0)}{x}=&\H{((0,1,0),0)}{x-1}\H{((1,0,1),0)}{1}\H{((1,1,0),0)}{z-1}\\
&-\H{((1,0,1),0)}{1}\H{((0,1,0),0),((1,1,0),0)}{z-1}\\
& -\H{((1,1,0),0)}{z-1}\H{((1,0,1),0),((-1,1,0),0)}{1}\\
&+\H{((0,1,0),0),((-1,1,0),0),((1,0,1),0)}{1}\\
&+\H{((1,1,0),0),((0,1,0),0),((2,2,1),0)}{z-1}\\
&+v\cdot i\pi\;\H{((1,0,1),0)}{1}\H{((1,1,0),0)}{z-1},
\end{align*}
%-----------------------------------------------------------------------------------------------------------------
where we set
%-----------------------------------------------------------------------------------------------------------------
\begin{eqnarray}
    v = \Biggl\{\begin{array}{rl}
        1, & \text{for adding $+i\ep$ to a singular letter}\\
        0, & \text{to compute Chauchy's principal value}\\
        -1, & \text{for adding $-i\ep$ to a singular letter}.
        \end{array}
\end{eqnarray}
%-----------------------------------------------------------------------------------------------------------------
Note that it might be necessary to perform the above steps several times to different poles.

The series expansion of a {\tt QL}-function $\H{\vec{m}}{z}$ with a letter 
%-----------------------------------------------------------------------------------------------------------------
\begin{eqnarray}
\frac{1}{x+a_i}
\label{eq:KP}
\end{eqnarray}
%-----------------------------------------------------------------------------------------------------------------
for $a_i \in \R\setminus \{0\}$ is 
convergent for $|z|<|a_i|$. Hence given a {\tt QL}-function $\H{\bf{m}}{z}$ with letters from $\mathfrak{A}_{KP}$, 
\ie with 
letters of the form (\ref{eq:KP}), 
we remove all poles $p$ at the real axis for which $|p|<|z|$ using the strategy mentioned above.
Now expanding the resulting functions about $z=0$ will lead to a convergent series at $z$. 

In the case that there are also quadratic forms present \ie letters of the form 
\begin{eqnarray}
\frac{1}{a_i+b_i\;x+ x^2}, \label{eq:QU}
\end{eqnarray}
we have to additionally treat the $\sqrt{|a_i|}$ as poles and apply the method mentioned 
above also with $p=\pm\sqrt{|a_i|}$ for $\sqrt{|a_i|}<|z|$. This is due to the fact that 
the series expansion of a {\tt QL}-function with a letter of the type (\ref{eq:QU}) 
is convergent for $|z|<\sqrt{|a_i|}$.

The option {\tt PrincipalValue $\rightarrow$ False} in {\tt QLEvaluate} allows to evaluate integrals sub--integrals containing 
poles yielding a 
complex result.
%-----------------------------------------------------------------------------------------------------------
\begin{eqnarray}
& & \int_0^1 dx \int_0^x dy \frac{1}{\left[\tfrac{3}{8} - x\right]} \frac{1}{\left[\tfrac{15}{8} - y\right]}
\rightarrow \lim_{\ep \rightarrow 0^+}\int_0^1 dx \int_0^x dy \frac{1}{\left[\tfrac{3}{8} - x + i \ep\right]} 
\frac{1}{\left[\tfrac{15}{8} - y\right]}
\nonumber\\ 
& = & -\zeta_2 + 4 \ln^2(2) + 2 \ln(2) \ln(3) + \ln^2(3) - 
\ln^2(5) - \left[2 \ln(2) + \ln(3) - \ln(5) \right] \ln(7) 
\nonumber\\ && 
+ \Li_2\left(-\frac{1}{4}\right) 
+ \Li_2\left(\frac{7}{12}\right)
+ i~\pi \left[2 \ln(2) - \ln(5) \right]
\nonumber\\ 
& = & {\tt QLEvaluate\left[QL\left[\{\{\{\tfrac{3}{8}, -1, 0\}, 0\}, 
\{\{\tfrac{15}{8}, -1, 0\}, 0\}\}, 1\right], 25, PrincipalValue \rightarrow False\right]} 
\nonumber\\
&\approx& -0.8205920210842043836307006 - 0.7010261415046584209879799~i.
\end{eqnarray}
%-----------------------------------------------------------------------------------------------------------
Here the regularization is performed in adding $+ i\ep$ to the denominator of the singular letters. 

The {\tt QL}--functions in the variable $z \in [0,1]$ can be Mellin transformed by
%-----------------------------------------------------------------------------------------------------------
\begin{eqnarray}
\Mvec[f(x)](N) = \int_0^1 dx x^N f(x).
\end{eqnarray}
%-----------------------------------------------------------------------------------------------------------
One obtains representations by harmonic sums \cite{Vermaseren:1998uu,Blumlein:1998if}
and (generalized) S--sums \cite{Ablinger:2013cf} at complex weights
%-----------------------------------------------------------------------------------------------------------
\begin{eqnarray}
S_{b,\vec{a}}(N) &=& \sum_{k=1}^N \frac{
({\rm sign}(b))^k
}{k^{|b|}} S_{\vec{a}}(k), S_\emptyset
= 1,~~
b, a_i \in \mathbb{Z} \backslash \{0\}
\\
S_{b,\vec{a}}(c,\vec{d},N) &=& \sum_{k=1}^N \frac{b^c}{k^c} S_{\{\vec{a}\}}(\vec{d},k), S_\emptyset 
= 1,~~
b, a_i \in \mathbb{C},~~c, d_i \in \mathbb{N} \backslash \{0\}.
\end{eqnarray}
%-----------------------------------------------------------------------------------------------------------
Instead of working with differential equations one may work with difference equations, which are solved by the package 
{\tt Sigma} \cite{SIG1,SIG2}. The ground field of the package needs then to contain the corresponding special 
constants arising in $\{\vec{c}\}$.

An example is the Mellin transformation of 
%-----------------------------------------------------------------------------------------------------------
\begin{eqnarray}
F_1(x) &=& {\tt QL\left[\left\{\{\{0, 1, 0\}, 0\right\}, \left\{\left\{\tfrac{1}{3}, 0, 1\right\}, 1\right\}, 
\left\{\left\{\tfrac{1}{3}, 0, 
1\}, 1\right\}\right\}, {\it x}\right]},
\end{eqnarray}
%-----------------------------------------------------------------------------------------------------------
which is given by
%-----------------------------------------------------------------------------------------------------------
\begin{eqnarray}
\label{ex:MVEC}
\Mvec[F_1(x)](N) &=& -
\frac{1}{(1+N)^4}
-\frac{9 T_1^2}{2 (1+N)^2}
-\frac{9 T_3}{1+N}
-\frac{9 T_4}{1+N}
+T_1 \Biggl(
        \frac{3}{(1+N)^3}
        +\frac{9 T_2}{1+N}
\Biggr)
\nonumber\\ &&
+\Biggl\{
        \Biggl[
                \frac{3 T_5}{2 (1+N)^2}
                + T_0 \Biggl(
                        \frac{1}{2 (1+N)^3}
                        -\frac{3 T_1}{2 (1+N)^2}
                        +\frac{S_1}{4 (1+N)^2}
                        -\frac{S_{-1}}{4 (1+N)^2}
                \Biggr)
\nonumber\\ &&
                + i \sqrt{3} T_1 \Biggl(
                        -\frac{1}{2 (1+N)^3}
                        -\frac{S_1}{4 (1+N)^2}
                        -\frac{S_{-1}}{4 (1+N)^2}
                        +\frac{S_1\big({{-i \sqrt{3}}}\big)}{2 (1+N)^2}
                \Biggr)
\nonumber\\ &&
                +\frac{3 i T_1^2 \sqrt{3}}{4 (1+N)^2}
                - \frac{i}{\sqrt{3}} \Biggl(
\frac{S_1\big({{-i \sqrt{3}}}\big)}{2 (1+N)^3}
                +\frac{S_{1,1}\big({{1,-i \sqrt{3}}}\big)}{4 (1+N)^2}
                +\frac{S_{1,1}\big({{-1,i \sqrt{3}}}\big)}{4 (1+N)^2} \Biggr)
        \Biggr] i^N
\nonumber\\ && 
        +\Biggl[
                \frac{3 T_5}{2 (1+N)^2}
                + T_0 \Biggl(
                        \frac{1}{2 (1+N)^3}
                        -\frac{3 T_1}{2 (1+N)^2}
                        +\frac{S_1}{4 (1+N)^2}
                        -\frac{S_{-1}}{4 (1+N)^2}
                \Biggr)
\nonumber\\ &&
                + i \sqrt{3} T_1
                 \Biggl(
                        \frac{1}{2 (1+N)^3}
                        +\frac{S_1}{4 (1+N)^2}
                        +\frac{S_{-1}}{4 (1+N)^2}
                        -\frac{S_1\big({{i \sqrt{3}}}\big)}{2 (1+N)^2}
                \Biggr)
\nonumber\\ &&
                -\frac{3 i T_1^2 \sqrt{3}}{4 (1+N)^2}
                + \frac{i}{\sqrt{3}} \Biggl(
                 \frac{    S_1\big({{i \sqrt{3}}}    \big)}{2 (1+N)^3}
                +\frac{S_{1,1}\big({{ 1, i \sqrt{3}}}\big)}{4 (1+N)^2}
                +\frac{S_{1,1}\big({{-1,-i \sqrt{3}}}\big)}{4 (1+N)^2}
\Biggr) 
\Biggr] 
\nonumber\\ && \times       
i^{3 N}
\Biggr\} 3^{-N/2},
\end{eqnarray}
%-----------------------------------------------------------------------------------------------------------
where we dropped the argument $N$ both for the harmonic and generalized harmonic sums.
The constants $T_i$ are iterated integrals at $x=1$ and are given by
%-----------------------------------------------------------------------------------------------------------
\begin{eqnarray}
T_0 &=& \frac{\pi}{3 \sqrt{3}}
\\
T_1 &=& \frac{\ln(2)}{3}
\\
T_2 &=& \frac{\zeta_2}{12} + \frac{\ln^2(3)}{24} + \frac{1}{12} \Li_2\left(-\frac{1}{3}\right)
\\
T_3 &=& \frac{1}{72} \zeta_3
-\frac{1}{27} \ln^3(2) + \frac{1}{36} \ln^2(2) \ln(3)  - 
 \frac{1}{36} \ln(2) \Li_2\left(\frac{1}{4}\right)
- \frac{1}{72} \Li_3\left(\frac{1}{4}\right)
\\
T_4 &=& \frac{1}{36} \zeta_2 \ln(2) - \frac{1}{36} \zeta_3 + \frac{1}{54} \ln^3(2) 
+ \frac{1}{36} \ln(2) \Li_2\left(\frac{1}{4} \right)
+ \frac{1}{36} \Li_3\left(\frac{1}{4}\right)
\\
T_5 &=& -\frac{1}{6 \sqrt{3}} {\sf Re}\Biggl[i \Li_2\left(\frac{1}{2} + i \frac{\sqrt{3}}{2}\right)\Biggr]
= 2 \zeta_2 - \frac{86}{33} \ln^2(3) + \frac{5}{33} \Li_2\left(-\frac{1}{3}\right),
\end{eqnarray}
%-----------------------------------------------------------------------------------------------------------
where $\Li_n(x)$ denotes the classical polylogarithm \cite{LEWIN} with the representation
%-----------------------------------------------------------------------------------------------------------
\begin{eqnarray}
\Li_n(x) = \sum_{k=1}^\infty \frac{x^k}{k^n},~~ x \in [-1,1], n \in \mathbb{N}.
\end{eqnarray}
%-----------------------------------------------------------------------------------------------------------
Not all of the above polylogarithms are independent, cf.~\cite{Ablinger:2014yaa,NOTES}, and the following relations hold,
%-----------------------------------------------------------------------------------------------------------
\begin{eqnarray}
\Li_2\left(\frac{1}{4}\right) &=& - \Li_2\left(-\frac{1}{3}\right) - 2 \ln^2(2) + 2 \ln(2) \ln(3) - \frac{1}{2} \ln^2(3)
\\
&=& 2 \Li_2\left(-\frac{1}{2}\right) + \zeta_2 - \ln^2(2)
\\
\Li_3\left(\frac{1}{4}\right) &=& 
\frac{7}{2} \zeta_3 - 2 \zeta_2 \ln(2) + \frac{2}{3} \ln^3(2) + 4 \Li_3\left(-\frac{1}{2}\right).
\end{eqnarray}
%-----------------------------------------------------------------------------------------------------------
The basis of contributing constants is here
%-----------------------------------------------------------------------------------------------------------
\begin{eqnarray}
\Biggl\{
\ln(2), \zeta_2, \zeta_3, \pi, \ln(3), \Li_2\left(-\frac{1}{2}\right),  \Li_3\left(-\frac{1}{2}\right) \Biggr\}. 
\end{eqnarray}
%-----------------------------------------------------------------------------------------------------------
The constants at position 1--3 are multiple zeta values \cite{Blumlein:2009cf}, at position 4 and 5 they are cyclotomic 
\cite{Ablinger:2011te}, and the ones at position  6 and 7 occurred already for square--root valued iterated integrals 
\cite{Ablinger:2014yaa}. The Mellin transformation (\ref{ex:MVEC}) is real.
%%%%%%%%%%%%%%%%%%%%%%%%%%%%%%%%%%%%%%%%%%%%%%%%%%%%%%%%%%%%%%%%%%%%%%%
\section{Physics Examples}
\label{sec:3}
%%%%%%%%%%%%%%%%%%%%%%%%%%%%%%%%%%%%%%%%%%%%%%%%%%%%%%%%%%%%%%%%%%%%%%%

\vspace*{1mm}
\noindent
As an application of the operations given in Section~\ref{sec:2} we calculate a few examples of particular iterated 
integrals, 
which have emerged in the calculation of inclusive Compton scattering cross sections
at next-to-leading order (NLO) \cite{Lee:2021iid} recently. The corresponding integrals 
were defined by
%-----------------------------------------------------------------------------------------------------------
\begin{eqnarray}
G_{b,\vec{a}}(x) = \int_0^x d w_b(x') G_{\vec{a}}(x'),
\label{eq:SCH}
\end{eqnarray}
%-----------------------------------------------------------------------------------------------------------
with 
%-----------------------------------------------------------------------------------------------------------
\begin{eqnarray}
d w_y(x) = \frac{y dx}{x},~~d w_a(x) = \frac{dx}{x-a},~~a \in \mathbb{R}, 
\end{eqnarray}
%-----------------------------------------------------------------------------------------------------------
where $y = \sqrt{x/(4+x)},~~x = s/m^2 - 1$, $s$ the cms energy and $m$ a mass, implying $y \in [0,1]$, and $y = 
y(x)$.

Iterative integrals obey shuffle algebras \cite{REUT,Borwein:1999js,Blumlein:2003gb} and are first 
reduced to a corresponding 
basis representation by exploiting the relations implied by (\ref{eq:shuf}). The formal iterated integrals 
(\ref{eq:SCH}) are given by
%-----------------------------------------------------------------------------------------------------------
\begin{eqnarray}
&& G_{0,y,-1}(x),~~~
   G_{y,-1,0}(x),~~~
   G_{y,0,-1}(x),~~~
   G_{y,-1,-1}(x),~~~
   G_{y,-2,-1}(x),~~~
   G_{-4,y,-1}(x), 
\nonumber\\
&& G_{y,-1}(x),~~~
   G_{y,y,-1}(x),
\label{eq:GL1}
\end{eqnarray}
%-----------------------------------------------------------------------------------------------------------
and $x = 4y^2/(1-y^2)$. 

We now cast these integrals into a root--free form. Furthermore, we choose a simple 
main argument,
to allow further iterated integration, needed in potential higher order calculations.
For the following representations we choose the variable
%-----------------------------------------------------------------------------------------------------------
\begin{eqnarray}
\label{eq:z}
z = \frac{1-y}{1+y} 
\end{eqnarray}
%-----------------------------------------------------------------------------------------------------------
for the main argument
and set $\HA_{\vec{a}}(z) \equiv \HA_{\vec{a}}$ for the harmonic polylogarithms and cyclotomic harmonic polylogarithms.

We now transform the integrals (\ref{eq:GL1}) into iterative integrals, see also \cite{AB1,AB2},
containing also letters generated by quadratic forms.
%-----------------------------------------------------------------------------------------------------------
\begin{eqnarray}
G_{0,y,-1}(x) &=&  
- \frac{4}{3} \zeta_3 
- \frac{1}{6} \HA^3  
+ \frac{2}{3} \zeta_2 \HA_1
+ \HA_0^2 \HA_1 
- \sqrt{3}\pi \HA_{\{3, 0\}, 0}(1) 
- 2 \HA_{0,0,1} 
- 2 \HA_{0, 0, \{6, 0\}} 
+ 2 \HA_{0, 0, \{6, 1\}} 
\nonumber\\ &&
- 2 \HA_{1, 0, \{6, 0\}} 
+ 4 \HA_{1, 0, \{6, 1\}}+ \frac{4z}{(1-z^2)^2} (\HA_{-1}(x) + \HA_{\{6, 0\}} - 2 \HA_{\{6, 1\}}) 
\nonumber\\ &&
+ \left[\frac{4z}{(1-z^2)^2} + \frac{1}{3} \zeta_2 + 2 \HA_{0,1} \right] \HA_0.
\label{eq:G1}
\end{eqnarray}
%-----------------------------------------------------------------------------------------------------------
Some of the iterated integrals in the r.h.s. in (\ref{eq:G1}) are obtained at main argument $x$, like $\HA_{-1}(x)$ 
and  $\HA_{0,-1}(x)$.
It appears to be useful to use the following relations
%-----------------------------------------------------------------------------------------------------------
\begin{eqnarray}
H_{-1}(x) &=&  -\HA_0 - H_{\{6, 0\}} + 2 \HA_{\{6, 1\}}.
\\
H_{-2,-1}(x) &=& 
\frac{1}{2} \HA_0^2  + \HA_{0, \{6, 0\}} 
- 2 \HA_{0, \{6, 1\}} - 2 \HA_{\{4, 1\}, 0} - 2 \HA_{\{4, 1\}, \{6, 0\}} 
+ 4 \HA_{\{4, 1\}, \{6, 1\}} - \frac{1}{2} \zeta_2
\\
H_{0,-1}(x) &=& \zeta_2 + \frac{1}{2} \HA_0^2 + 2 \HA_0 \HA_1 - 2 \HA_{0,1}
+ 
 \HA_{0, \{6, 0\}} - 2 \HA_{0, \{6, 1\}} + 2 \HA_{1, \{6, 0\}} - 
 4 \HA_{1, \{6, 1\}}.
\end{eqnarray}
%-----------------------------------------------------------------------------------------------------------
which yields
%-----------------------------------------------------------------------------------------------------------
\begin{eqnarray}
\label{eq:CYC1}
G_{0,y,-1}(x) &=&  
- \frac{4}{3} \zeta_3 
- \frac{1}{6} \HA^3_0  
+ \frac{2}{3} \zeta_2 \HA_1
+ \HA_0^2 \HA_1 
- \sqrt{3}\pi \HA_{\{3, 0\}, 0}(1)
- 2 \HA_{0,0,1} 
- 2 \HA_{0, 0, \{6, 0\}} 
+ 2 \HA_{0, 0, \{6, 1\}} 
\nonumber\\ &&
- 2 \HA_{1, 0, \{6, 0\}} 
+ 4 \HA_{1, 0, \{6, 1\}}
+ \left[\frac{1}{3} \zeta_2 + 2 \HA_{0,1} \right] \HA_0.
\end{eqnarray}

For the other integrals the following results are obtained,
%-----------------------------------------------------------------------------------------------------------
\begin{eqnarray}
 G_{y,-1,0}(x) &=& \frac{2}{3} \zeta_3 + \zeta_2 \HA_0 - \frac{1}{6} \HA_0^3  - 2 \HA_{0,0,1}
- \HA_{0, \{6, 0\}, 0} - 2 \HA_{0, \{6, 0\}, 1} + 2 \HA_{0, \{6, 1\}, 0} 
\nonumber\\ &&
+ 4 \HA_{0, \{6, 1\}, 1}
\\ 
   G_{y,0,-1}(x) &=& \frac{2}{3} \zeta_3 - \zeta_2  \HA_0 -\frac{1}{6} \HA_0^3
-2 \HA_0 \HA_{0,1} + \sqrt{3} \pi \HA_{\{3,0\},0}(1) + 4 \HA_{0,0,1} - \HA_{0,0,\{6,0\}}
\nonumber\\ &&
+2 \HA_{0,0,\{6,1\}} 
- 2 \HA_{0,1,\{6,0\}} + 4 \HA_{0,1,\{6,1\}}
\\
G_{y,-1,-1}(x) &=& \frac{5}{9} \zeta_3 -\frac{1}{6} \HA_0^3 + \frac{\pi}{\sqrt{3}} \HA_{\{3,0\},0}(1)
-\HA_{0,0,\{6,0\}}
+2 \HA_{0,0,\{6,1\}}
-\HA_{0,\{6,0\},0}
-\HA_{0,\{6,0\},\{6,0\}}
\nonumber\\ &&
+2 \HA_{0,\{6,0\},\{6,1\}}
+2 \HA_{0,\{6,1\},0}+2 \HA_{0,\{6,1\},\{6,0\}}
-4 \HA_{0,\{6,1\},\{6,1\}}
\\
G_{y,-2,-1}(x) &=& \frac{1}{12} \zeta_3 + \frac{1}{2} 
\zeta_2 \HA_0
- \frac{1}{6} \HA_0^3  - \frac{1}{4} \sqrt{3} \pi \HA_{\{3,0\},0}(1)
- \HA_{0, 0, \{6, 0\}} 
\nonumber\\ &&
+ 2 \HA_{0, 0, \{6, 1\}} + 2 \HA_{0, \{4, 1\}, 0} 
+ 2 \HA_{0, \{4, 1\}, \{6, 0\}} - 4 \HA_{0, \{4, 1\}, \{6, 1\}}
\\ 
G_{-4,y,-1}(x) &=& \frac{16}{9} \zeta_3 - \frac{1}{6} \HA_0^3 + \HA_{-1} \left(-\frac{2}{3} \zeta_2 + \HA_0^2 \right) + 
 \HA_0 \left(\frac{1}{2} \zeta_2 - 2 \HA_{0,-1}\right) + \frac{2 \pi}{\sqrt{3}} \HA_{\{3, 0\}, 0}(1) 
\nonumber\\ && 
+ 2 \HA_{-1, 0, \{6, 0\}} - 4 \HA_{-1, 0, \{6, 1\}} + 2 \HA_{0, 0, -1} - 
 \HA_{0, 0, \{6, 0\}} + 2 \HA_{0, 0, \{6, 1\}}
\\
G_{y,-1}(x) &=& -\frac{1}{3} \zeta_2 + \frac{1}{2} \HA_0^2  + \HA_{0, \{6, 0\}} - 2 \HA_{0, \{6, 1\}}
\\
G_{y,y,-1}(x)  &=& \frac{2}{3} \zeta_3 + \frac{1}{3} \zeta_2 \HA_0 - \frac{1}{6} \HA_0^3 - \HA_{0, 0, \{6, 0\}} 
+  2 \HA_{0, 0, \{6, 1\}},
\label{eq:CYC2}
\end{eqnarray}
%-----------------------------------------------------------------------------------------------------------
where 
%-----------------------------------------------------------------------------------------------------------
\begin{eqnarray}
H_{\{3,0\},0}(1) = - \frac{2}{\sqrt{3}} {\rm Cl}_2\left(\frac{2}{3} \pi\right) = \frac{2}{9} \left[4 \zeta_2 - 
\psi'\left(\frac{1}{3}\right) \right]
\end{eqnarray}
%-----------------------------------------------------------------------------------------------------------
is another cyclotomic constant and ${\rm Cl}_2$ denotes a Clausen functions with
%-----------------------------------------------------------------------------------------------------------
\begin{eqnarray}
{\rm Cl}_2(x) = \sum_{k=1}^\infty \frac{\sin(kx)}{k^2}
\end{eqnarray}
%-----------------------------------------------------------------------------------------------------------
and $\psi'$ denotes the first derivative of the di-gamma function.

The above functions contain letters of the usual harmonic polylogarithms and those of cyclotomy {\sf c = 3,4} and 
{\sf 6} \cite{Ablinger:2011te}, to which the notation in (\ref{eq:CYC1}--\ref{eq:CYC2}) corresponds. This is the case 
because of the choice of the variable 
$z$, which is possible in the 
univariate case, although one does not know the optimal choice  a priori. The cyclotomic letters 
are characterized 
by two indices \cite{Ablinger:2011te} as e.g. 
%-----------------------------------------------------------------------------------------------------------
\begin{eqnarray}
f_{\{5,3\}}(x) = \frac{x^3}{1 + x + x^2 + x^3 + x^4},
\end{eqnarray}
%-----------------------------------------------------------------------------------------------------------
where the second index is smaller than the degree of the denominator polynomial.

Choosing the variable $y$ instead as the main argument, one transforms the above expressions
using Eq.~(\ref{eq:MOB1}). As a consequence the additional letters
%-----------------------------------------------------------------------------------------------------------
\begin{eqnarray}
\left\{ \frac{1}{1+ 3 x^2}, \frac{x}{1+ 3 x^2}\right\}
\end{eqnarray}
%-----------------------------------------------------------------------------------------------------------
are needed, which are not cyclotomic. As an example one obtains
%-----------------------------------------------------------------------------------------------------------
\begin{eqnarray}
G_{y,-1}(x) &=& -\frac{1}{2} \HA_{-1}^2(y) 
- \HA_{-1}(y)  \HA_1(y) 
+ \frac{1}{2} \HA_1^2(y)
+ 2 \HA_{-1, 1}(y) 
\nonumber\\ &&
-2 {\tt QL\left[\{\{\{-1, 1, 0\}, 0\}, \{\{\tfrac{1}{3}, 0, 1\}, 1\}\}, {\it y}\right]} + 
 2 {\tt QL\left[\{\{\{1, 1, 0\}, 0\}, \{\{\tfrac{1}{3}, 0, 1\}, 1\}\}, {\it y}\right]},
\nonumber\\
\end{eqnarray}
%-----------------------------------------------------------------------------------------------------------
and similar in the case of the other integrals.
In general also {\tt QL}--functions at argument $x=1$ will appear, inducing new constants. This set 
can be reduced to the algebraic basis by shuffle and stuffle relations \cite{Borwein:1999js}. Depending on the defining 
constants of the contributing quadratic forms, further sets of relations may be present. Often 
particular algebraic numbers occur in this context.
%%%%%%%%%%%%%%%%%%%%%%%%%%%%%%%%%%%%%%%%%%%%%%%%%%%%%%%%%%%%%%%%%%%%%%%
\section{Conclusions}
\label{sec:4}
%%%%%%%%%%%%%%%%%%%%%%%%%%%%%%%%%%%%%%%%%%%%%%%%%%%%%%%%%%%%%%%%%%%%%%%

\vspace*{1mm}
\noindent
In the analytic calculation of Feynman diagrams hierarchies of function spaces and algebras emerge both in the
$z$ space and $N$ space representation consisting out of iterative integrals or nested sums of different kind. The 
simplest structures are 
harmonic 
polylogarithms, followed by Kummer--Poincar\'{e} iterated integrals and  cyclotomic integrals, and also iterated 
integrals over square--root valued letters and the associated sums and special constants. 
Here we consider a real extension of 
the Kummer--Poincar\'{e} iterated 
integrals, allowing also for letters generated by general real quadratic forms without real factorization.
Here the range of constants is not limited to $c_i \in \mathbb{Q}$, but general real numbers are allowed, which 
are usually implied by the values of different masses and virtualities in the processes to be considered.
Quantities of this kind appear in higher order and multi-leg calculations. Since real representations have sometimes 
advantages compared to complex representations, we provide algorithms to build the associated algebra to a set 
of these letters, their basis representation, different mappings of the main argument, including analytic continuation 
in the case of the presence of cuts. Finally, also the expressions can be evaluated numerically. The different commands 
in {\tt HarmonicSums} to provide these operations are described and illustrated by examples. We have applied the
corresponding mappings to a class of functions which have emerged recently in the NLO calculation of the inclusive
Compton cross section. In viewing physics results within this class, it is for structural reasons also interesting to 
see whether the result can be expressed by functions out of a particular function space. In the case of 
Ref.~\cite{Lee:2021iid} it turns out to be the space of cyclotomic harmonic polylogarithms if the final expression is
written using the variable $z$, (\ref{eq:z}).

The different commands to treat iterated integrals of the {\tt QL}--type are implemented in the 
package {\tt HarmonicSums}, which is available from {\tt https://risc.jku.at/sw/harmonicsums/}. As well we attach the 
ancillary file {\tt QuadraticLetters.nb} to this paper.

\vspace{5mm}\noindent
{\bf Acknowledgment.}~
We thank P.~Marquard and S.~Weinzierl for a discussion. This project has received funding from the European Union's 
Horizon 2020 research 
and innovation programme under the Marie Sk\l{}odowska--Curie grant agreement No. 764850, SAGEX and from the Austrian Science 
Fund (FWF) grant SFB F50 (F5009-N15). 
%--------------------------------------------------------------------------------------------------

%-----------------------------------------------------------------------------------
\end{document}